\documentclass[12pt,a4paper]{article}

\usepackage{epsfig}
\usepackage{amssymb}

\def\be{\begin{equation}}
\def\ee{\end{equation}}

\begin{document}

\title{
\begin{flushright} \small \rm
  hep-ph/0007107 \\ 
  DESY 99--181 \\ 
  FTUV/00-08 \\ 
  IFIC/00-08 \\
  UWThPh-2000/24 \\[1cm]
\end{flushright} 
Reconstructing Supersymmetric Theories \\ at High Energy Scales}
\author{G.A.~Blair$^{a,b}$, W.~Porod$^{c,b}$, and P.M.~Zerwas$^{b,d}$\\[0.5cm] 
\small
$^a$ Royal Holloway and Bedford New College, University of London, London,
 UK\\ \small
$^b$ Deutsches Elektron--Synchrotron DESY, D-22603 Hamburg, Germany \\ \small
$^c$ Inst.~de F\'\i sica Corpuscular (IFIC), CSIC, E-46071--Val\`encia,
      Spain \\ \small
$^d$Inst.~f.~Theor.~Physik, Universit\"at, A-1090 Vienna, Austria 
}
\date{\today}
\maketitle

\begin{abstract}
We have studied the reconstruction of supersymmetric
theories at high scales by evolving the fundamental parameters
from the electroweak scale upwards. Universal minimal supergravity and
gauge mediated supersymmetry breaking have been taken as representative
alternatives. Pseudo-fixed point structures require the low--energy
boundary values to be measured with high precision.
\end{abstract}

{\bf 1.} Supersymmetric theories in which fermionic and bosonic
particles are assigned to common multiplets, allow stable extrapolations
to high energy scales of order $M_U \simeq 2 \cdot 10^{16}$~GeV where
the electroweak and the strong couplings are expected to unify \cite{GUT}.
Since supersymmetry is not an exact symmetry,
a variety of breaking
mechanisms have been proposed, based on rather different physical
ideas. Among these schemes are supergravity theories \cite{sugra} and
gauge mediated supersymmetry breaking \cite{gmsb}.
The scales at which these mechanisms become effective, extend from the
grand unification scale near $10^{16}$~GeV down to scales as low as order
1 TeV. First indications 
about which of the scenarios could be realized in Nature,
may be derived from the mass spectrum once supersymmetric particles are 
observed experimentally \cite{comparison}. Moreover, dynamical signatures 
can be exploited for gauge mediated supersymmetry breaking, for example,
such as delayed photon decays of the lightest neutralino
or stau state \cite{blair}. 

In this note we address the extent to which the structure
of supersymmetric theories at high scales can be reconstructed directly
from future precision measurements of the properties of supersymmetric
particles. Since the structure of the theory at the high scale cannot
be assumed known
{\it a priori}, top-down approaches may not lead to valid
conclusions in general
so that bottom-up approaches provide the only proper method. 
While top-down
approaches have been discussed frequently in the literature, see 
e.g.~Refs.~\cite{blair,LC,LHC,blair2}, 
the direct reconstruction of the supersymmetric theory at the high scale,
being much
more difficult in practice, has not widely been addressed before. 
Theoretical elements in the context of fixed-point structures have been
discussed in Refs.~\cite{Allanach}.
The analysis in this note is phenomenological in nature,
based on the experimental
accuracies expected in the supersymmetric particle sector
at the proton--proton collider LHC  
and combined with
expectations from a future $e^+ e^-$ linear collider LC.

As paradigm we will choose minimal supergravity (mSUGRA).
The universal set of soft supersymmetry breaking parameters in this theory is
generated near the Planck scale 
where supersymmetry breaking is
mediated by gravity from a hidden sector \cite{nilles}.
Deviations from the universal values of the gaugino and scalar masses
may be induced by the evolution down to the grand-unification scale (GUT)
of the gauge couplings \cite{nir},
or by contributions from non-singlet F-terms; the deviations may even be
dramatic 
in superstring models (cf.~Ref.~\cite{drees} for details). 
Since the pattern of the mass terms
may therefore not be regular at the GUT scale itself, 
the bottom-up approach is 
needed to uncover these more complicated structures.

We will confront the mSUGRA extrapolation with the alternative gauge
mediated supersymmetry breaking (GMSB), characterized by a messenger
scale $M_m$ in the
range between $\sim 10$ TeV and $\sim 10^5$~TeV. 
In this scenario the mass parameters
of particles carrying the same gauge quantum numbers squared are universal.
The regularity for scalar masses would be observed at the scale $M_m$
while the gaugino mass parameters should unify at 1-loop order 
at the GUT scale $M_U$ as before. 

{\bf 2.}  
The extrapolation from the electroweak scale to the GUT scale 
in the mSUGRA scenario
is based on 
the supersymmetric renormalization group equations \cite{RGE1}. To
leading order, the gauge couplings and the gaugino and scalar mass 
parameters of soft supersymmetry breaking
depend on the evolution coefficients,
\be
Z_i = \left[ 1 + b_i {\alpha_U \over 4 \pi}
             \log\left({M_U \over M_Z}\right)^2 \right]^{-1} \, \, ,
\ee
with $b[SU_3, SU_2, U_1] = -3, \, 1, \, 33 / 5$;
the scalar mass parameters depend 
also on the Yukawa couplings $h_t$, $h_b$, $h_\tau$
of the
top quark, bottom quark and $\tau$ lepton. Denoting the unified coupling at 
the GUT scale $M_U$ by $\alpha_U$, the universal gaugino mass by $M_{1/2}$,
the universal sfermion and Higgs mass parameter by $M_0$,  and the universal
trilinear coupling by $A_0$, the renormalization group 
equations lead to the following relations for the low-scale parameters
\cite{solutions}: 
\begin{eqnarray}
\mathrm{gauge \,\, couplings}  & : & \alpha_i = Z_i \, \alpha_U \\
\label{eq:gaugino}
\mathrm{gaugino \,\, mass \,\, parameters} & : & M_i = Z_i \, M_{1/2} \\
\mathrm{scalar \,\, mass \,\, parameters} & : &
\label{eq:squark} 
M^2_j = M^2_0 + c_j M^2_{1/2}
       + {\textstyle \sum_{\beta=1,2}} c'_{j \beta} \Delta M^2_\beta \\
\mathrm{trilinear \,\, couplings} & : & A_k = d_k A_0  \, + d'_k M_{1/2} 
\end{eqnarray}
The coefficients $c_j$ [$j=L_l, E_l, Q_l, U_l, D_l, H_{1,2}$; $l=1,2,3$] 
for the slepton and squark doublets/singlets of generation $l$ and for the 
Higgs doublets
are linear combinations of the evolution
coefficients $Z_i$, the coefficients $c'_{j \beta}$ are of order unity. 
The shifts $\Delta M^2_\beta$ are nearly zero for the first two families of 
sfermions but they can be rather large for the third family and the 
Higgs mass
parameters, depending on the coefficients $Z_i$, 
the universal parameters $M^2_0$, $M_{1/2}$ and $A_0$,
and on the Yukawa couplings $h_t$, $h_b$, $h_\tau$.
The coefficients $d_k$ of the trilinear
couplings $A_k$ [$k=t,b,\tau$]  
depend on the corresponding Yukawa couplings 
and are approximately unity for the
first two generations while being O($10^{-1}$) 
and smaller if the Yukawa couplings are
large; the coefficients $d'_k$, depending on gauge 
and Yukawa couplings, are of order unity.

In the present analysis the evolution equations have been solved to
two--loop order \cite{RGE2} and threshold effects have been
incorporated at the low scale \cite{bagger}.  We have checked that the
points under study are compatible with $b \to s \gamma$ \cite{CLEO}
and the $\rho$-parameter \cite{Drees90}.  
The mSUGRA point we have analyzed in detail, is
characterized by the following parameters: $M_{1/2} = 190$~GeV, $M_0 =
200$~GeV, $A_0$ = 550~GeV, $\tan \beta = 30$, and $\mathrm{sign}(\mu)
= -$. The modulus of $\mu$ is calculated from the requirement of
radiative electroweak symmetry breaking.

The initial ``experimental'' values, 
are generated by evolving the universal parameters 
down to the electroweak scale according to standard
procedures \cite{aarason,bagger}. These parameters define the
experimental observables, including the supersymmetric particle masses
and production cross sections, which are endowed with errors as
derived from detailed experimental simulations of future LHC
\cite{LHC} and LC measurements \cite{blair2}. The LC errors on the
masses of sleptons and charginos/neutralinos 
are derived from the reconstruction of supersymmetric particles
in the continuum and from scanning the threshold regions; the
threshold analysis provides in general the most accurate value. 
The analysis of the entire particle spectrum requires LC energies 
up to 1 TeV and an integrated luminosity of about 1 ab$^{-1}$.
The errors given in  Ref.\cite{blair2} are scaled
in proportion to the masses of the spectrum. Moreover, 
they are inflated conservatively for particles that
decay predominantly to $\tau$ channels, according to typical reconstruction
efficiencies such as given in Ref.\cite{Nojiri96}.
Typical examples are shown in Table~\ref{tab:masserrors}.
The LC errors on the squark masses (see {\it e.g.} Ref.\cite{Feng})
are set to an average value of 3 GeV;
varying this error within a factor two does not change the conclusions
significantly since the measurement of the cross sections
provides the maximal sensitivity in this sector.
For the cross sections we use purely statistical errors, 
assuming a conservative
reconstruction efficiency of 20\%.
Parameter combinations from the 
fits to the spectrum and the cross sections which lead 
to charge and/or color breaking minima \cite{Casas},
are not accepted.

\begin{figure}
\setlength{\unitlength}{1mm}
\begin{center}
\begin{picture}(150,195)
\put(-18,-14){\mbox{\epsfig{figure=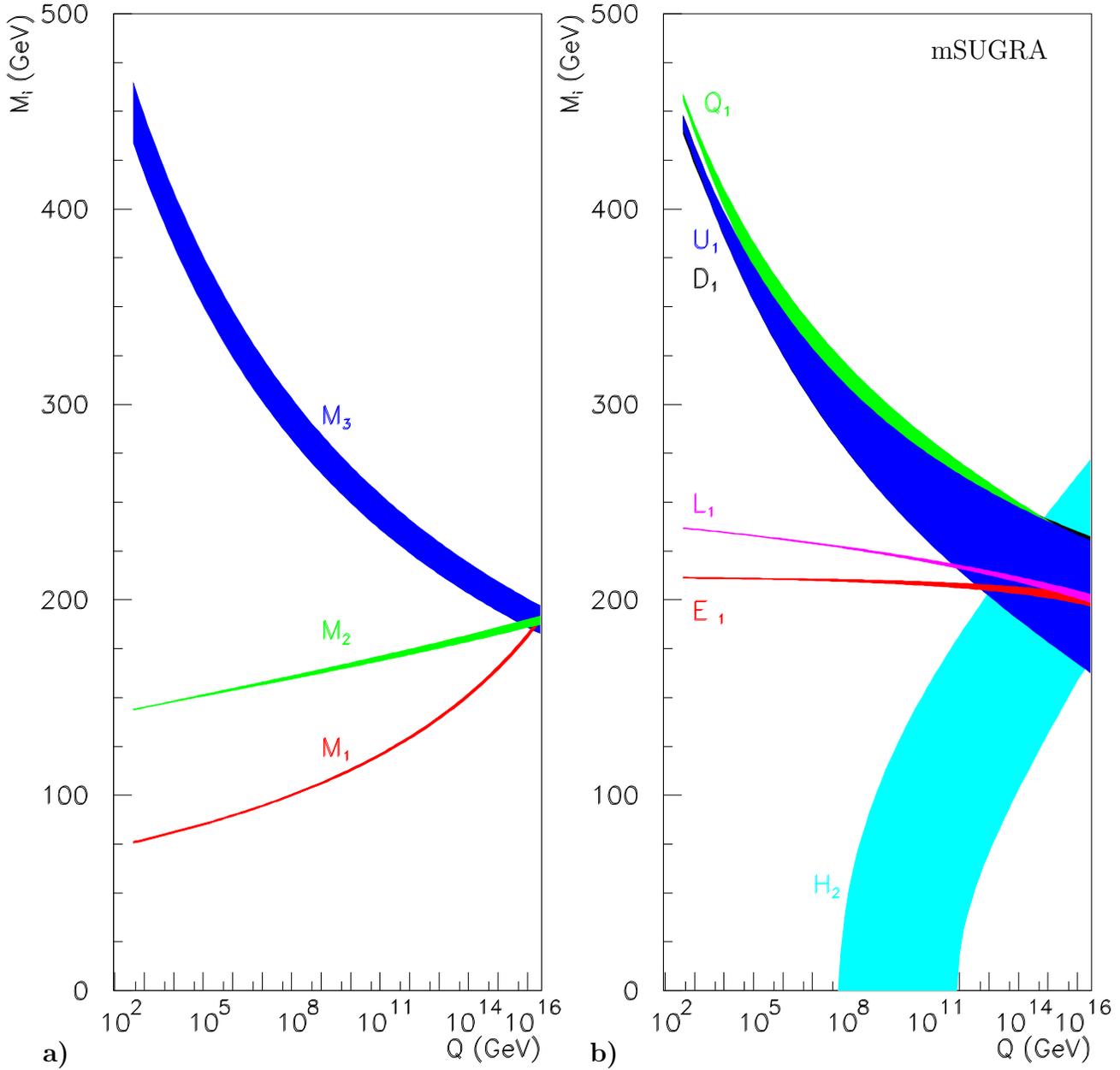,height=19.0cm,width=19.cm}}}
\put(-10,-6){\mbox{\bf a)}}
\put(75,-6){\mbox{\bf b)}}
\put(128,150){\mbox{mSUGRA}}
\end{picture}
\end{center}
\caption{{\bf mSUGRA:} {\it Evolution 
of (a) gaugino and (b) sfermion mass parameters in the
bottom--up approach. The mSUGRA point probed is characterized by the
parameters $M_0 = 200$~GeV, 
$M_{1/2} = 190$~GeV, $A_0$ = 550~GeV, $\tan \beta = 30$, 
and $\mathrm{sign}(\mu) = (-)$.
[The widths of the bands indicate the 95\% CL.]
}} 
\label{fig:sugra}
\end{figure}

These observables are interpreted as the experimental input values for
the evolution of the mass parameters in the bottom-up
approach to the grand unification scale. 
The results for the evolution of the mass parameters to the
GUT scale $M_U$  are shown in Fig.~\ref{fig:sugra}.
The left-hand side (a) of the figure presents the evolution of the gaugino
parameters $M_i$ which apparently is under excellent control, 
as is the extrapolation
of the slepton mass parameter in Fig.~\ref{fig:sugra}(b). The accuracy
deteriorates for the squark mass parameters and for the Higgs mass parameter
$M_{H_2}$.
The origin of the differences between the errors for slepton, squark, and
Higgs mass parameters can be traced back to the size of the coefficients
in Eqs.~(\ref{eq:squark}) for which typical examples read as
follows:
\begin{eqnarray}
M^2_{\tilde L_{1}} &\simeq& M_0^{2} + 0.52 M^2_{1/2} \\
M^2_{\tilde Q_{1}} &\simeq& M_0^{2} + 6.7 M^2_{1/2}  \\
M^2_{\tilde H_2} &\simeq&  -0.18 M_0^{2} - 2.2 M^2_{1/2}
           - 0.35 A_0 M_{1/2} - 0.08 A^2_0 
\end{eqnarray}
While the coefficients for sleptons are of order unity, the coefficient $c_j$
for squarks grows very large,  $c_j \simeq 6.7$, so that small errors
in $M^2_{1/2}$ are magnified by nearly an order of magnitude in the solution
for $M_0$. By close inspection of Eq.~(\ref{eq:squark}) for the Higgs mass
parameter it turns out that 
the formally leading 
$M^2_0$ part is nearly canceled by the $M^2_0$ part
of $c'_{j,\beta} \Delta M_\beta^2$. Inverting Eq.~(\ref{eq:squark}) for
$M^2_0$ therefore gives rise to large errors in the Higgs case.
A representative
set of mass values and the associated errors, 
as evolving from the electroweak scale to $M_U$, is
presented in Table~\ref{tab:parvalues}. 
The accuracy improves considerably if the
LHC measurements are complemented by the high--precision LC measurements. 
Extracting the trilinear parameters $A_k$ is difficult and
more refined analyses
based on sfermion cross sections and Higgs and/or sfermion decays are
necessary to determine these parameters more accurately. Moreover, the
$A_t$ coupling, the best measured coupling among the $A_k$ parameters,
shows a pseudo--fixed point behavior \cite{Allanach} since $d_t \simeq 0.2$
is small compared to $d'_t \simeq 2$. All other trilinear couplings have
only a weak impact on physical observables so that 
large experimental errors are expected. As a result, the
fundamental parameter $A_0$ cannot be determined as precisely as the other
parameters at the GUT scale.  

It is apparent from this discussion that the errors in extracting the squark
mass parameter $M_0$ depend strongly on whether $M_0$ is larger than $M_{1/2}$,
the case studied above, or whether $M_0$ is smaller than $M_{1/2}$. 
As an example in the latter case, the large Yukawa couplings of the third
generation can enhance the
pseudo--fixed point behavior, leading to large errors for $M_0$ in the
third generation.
                
Inspecting Fig.~\ref{fig:sugra}(b) leads to the conclusion that the 
top-down approach eventually may generate an incomplete picture. 
{\it Global} fits based on mSUGRA without allowing for deviations
from universality, are dominated by $M_{1,2}$ and the slepton mass
parameters due to the pseudo-fixed point behavior of the squark mass
parameters.  Therefore, the structure of the theory in the squark sector
is not scrutinized stringently at the unification scale
in the top-down approach. 
By contrast, the bottom-up approach demonstrates very clearly the extent
to which the theory can be tested at the high scale.

\begin{figure}
\setlength{\unitlength}{1mm}
\begin{center}
\begin{picture}(150,195)
\put(23,-14){\mbox{\epsfig{figure=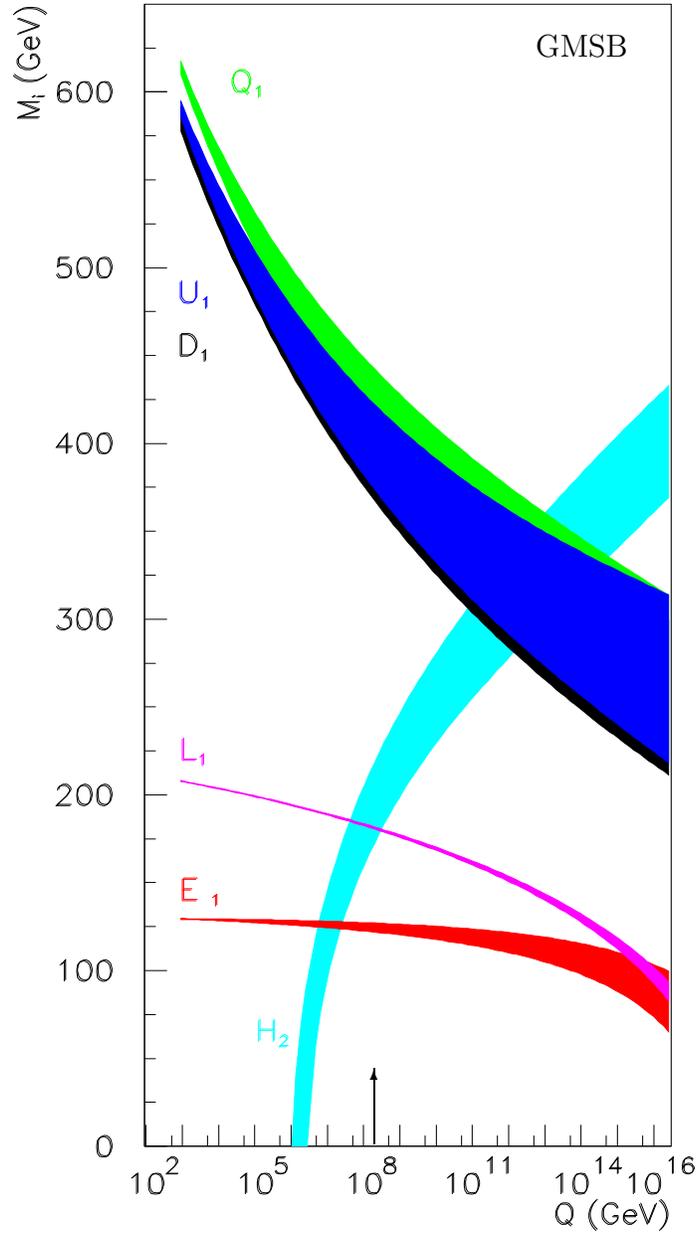,height=19.0cm,width=20.cm}}}
\put(73.5,5.3){\vector(0,1){10}}
\put(95,150){\mbox{GMSB}}
\end{picture}
\end{center}
\caption{{\bf GMSB:} {\it Evolution of sfermion 
mass parameters in the bottom--up approach. The GMSB point has been 
chosen as $M_m = 2 \cdot 10^5$~TeV, 
$\Lambda = 28$~TeV, $N_5 = 3$, $\tan \beta = 30$, 
and $\mathrm{sign}(\mu) = (-)$. 
[The widths of the bands indicate the 95\% CL.]
}} 
\label{fig:gmsb}
\end{figure}

{\bf 3.} To confront the mSUGRA analysis with an alternative
scenario, the analysis has been repeated at energies up to 1.5 TeV for gauge 
mediated supersymmetry breaking GMSB.
Regularity among particles carrying the
same gauge quantum numbers squared, should in this scenario be observed in 
the evolution of mass parameters at the messenger scale. 
The evolution of the 
sfermion mass parameters of the first/second
generation and the Higgs mass parameter $M_{H_2}$
is presented in Fig.~\ref{fig:gmsb}. It is obvious that $M_{H_2}$
approaches the 
mass parameter for the
left-chiral sleptons at the GMSB scale. Moreover, the
figure demonstrates clearly that GMSB will not be confused with 
the mSUGRA scenario as no more regularity can be observed
at the GUT scale $M_U$.

{\bf 4.} In summary.
The model--independent reconstruction of the fundamental
supersymmetric theory at the high scale, the grand unification scale $M_U$ in
supergravity or the intermediate scale $M_m$ in gauge mediated supersymmetry 
breaking, appears feasible. Regular patterns can be observed by evolving the 
gaugino and scalar mass parameters from the measured values at the electroweak
scale to the high scales. The accuracy is significantly improved if, in 
addition to the LHC input values, high--precision LC values are also included.
The future experimental input from LC is  particularly important if the
universality at the GUT scale is (slightly) broken. Precision
data are therefore essential for stable
extrapolations to high energy scales.

\section*{Acknowledgments}

Thanks for discussions go to M.~Peskin and D.~Pierce.
W.P.~and G.B.~are grateful to DESY for the warm hospitality during their stay
in Hamburg, P.M.Z. to the Institut f\"ur Theoretische Physik of Vienna
University. 
W.P.~is supported by the Spanish 'Ministerio de Educacion y Cultura' under 
the contract SB97-BU0475382.

\newpage

\begin{table}
\caption[]{\it Representative experimental mass errors used in the
fits to the mass spectra (see the text for details).}
\label{tab:masserrors}
\begin{center}
\begin{tabular}{|c||c|c|c|}
\hline
Particle         & M(GeV) & \multicolumn{2}{|c|}{$\Delta$ M(GeV)}\\
                 & Mass   & LHC & LHC+LC   \\\hline \hline
$h^0$            & 109    & 0.2 & 0.05    \\
$A^0$            & 191    &  3  & 1.5     \\ \hline
$\chi^+_1$            & 133    &  3  & 0.11    \\
$\chi^0_1$            & 72.6   &  3  & 0.15    \\ \hline
$\tilde{\nu_e}$  & 233    &  3  & 0.1    \\
$\tilde{e_1}$    & 217    &  3  & 0.15    \\
$\tilde{\nu_\tau}$& 214   &  3  & 0.8    \\
$\tilde{\tau_1}$& 154     &  3  & 0.7  \\ \hline
$\tilde{u_1}$   & 466     &  10 & 3   \\
$\tilde{t_1}$   & 377     &  10 & 3   \\ \hline
$\tilde{g}$     & 470     &  10 & 10   \\
\hline
\end{tabular}
\end{center}
\end{table}

\begin{table}
\caption[]{\it Representative mass parameters as determined at the electroweak 
scale and evolved to the GUT scale; based on LHC (left--hand side),
and LC simulations (right--hand side). $L_{1,3}$, $Q_{1,3}$ are the slepton and
squark isodoublet parameters of 
the first and third family; the minus sign ($-$)
in front of $M_{H_2}$ refers to the negative value of  $M^2_{H_2}$ at the
electroweak scale. [The errors quoted correspond to 1$\sigma$.]}
\label{tab:parvalues}
\begin{center}
\begin{tabular}{|c||c|c||c|c|}
\hline
 & \multicolumn{2}{|c||}{LHC}  & \multicolumn{2}{|c|}{LC} \\ \hline
 & exp.~input & GUT value & exp.~input & GUT value \\ \hline  \hline  
 $M_1 $ & 75.6 $\pm$ 3.2      & 189.6 $\pm$ 7.6
        & 75.6  $\pm$  0.2  & 189.6 $\pm$ 0.7 \\
 $M_2 $ & 143.6 $\pm$ 3.1     & 190.6 $\pm$ 3.8 
        & 143.6 $\pm$ 0.2   & 189.4 $\pm$ 0.9\\
 $M_3 $ & 452.3 $\pm$ 11.9    & 190.1 $\pm$ 5.7 
        & 452.3 $\pm$ 9  & 190.0 $\pm$ 4.2         \\ \hline 
 $M_{L_1} $ & 236.8 $\pm$ 2.1  & 200.6 $\pm$ 6.9
            & 236.8  $\pm$  0.1 & 200.5 $\pm$ 0.9\\
 $M_{Q_1} $ & 459.6 $\pm$ 7.4 & 200.7 $\pm$ 30.5 
            & 459.7 $\pm$ 0.6 & 200 $\pm$ 18 \\ \hline 
 $M_{L_3} $ & 218.6$\pm$ 2.8 & 199.5 $\pm$ 12.3 
            & 218.6 $\pm$ 0.6 & 196.5 $\pm$ 7.2\\
 $M_{Q_3} $ & 392 $\pm$ 45 & 192 $\pm$ 251 
            & 391.2 $\pm$ 1.0 & 233 $\pm$ 46\\ \hline 
 $M_{H_1} $ & 132.4 $\pm$12  & 361 $\pm$ 324 
            & 132.4 $\pm$ 1.5  & 224 $\pm$ 90\\
 $|M_{H_2}|$ & (-)251.9 $\pm$2.2 & 279 $\pm$ 98
             & (-)251.9$\pm$0.2 & 211 $\pm$ 27\\ \hline 
 $A_\tau $ & 101 $\pm$ 2590  & 210 $\pm$ 432
           &100 $\pm$ 92 & 319 $\pm$ 340\\
 $A_b $ & -125 $\pm$ 3920  & 806 $\pm$ 1292
        &-126$\pm$ 286 & 129 $\pm$ 571\\
 $A_t $ & -186 $\pm$ 39  & 608 $\pm$ 169
         &-186.3 $\pm$3.2 & 505 $\pm$ 81\\ \hline
\end{tabular}
\end{center}
\end{table}

\end{document}